\documentclass[a4paper,aps,prl,floatfix,reprint,superscriptaddress,showpacs,twocolumn]{revtex4-1}
\usepackage[latin9]{inputenc}
\usepackage{color}
\usepackage{textcomp}
\usepackage{mathrsfs}
\usepackage{amsmath}
\usepackage{amssymb}
\usepackage{graphicx}
\usepackage[unicode=true,
 bookmarks=false,
 breaklinks=false,pdfborder={0 0 1},backref=false,colorlinks=true]
 {hyperref}
\hypersetup{
 letterpaper=true,pdfstartview=FitV,linkcolor=blue,citecolor=blue,urlcolor=blue}

\makeatletter

\pdfpageheight\paperheight
\pdfpagewidth\paperwidth

\pdfoutput=1
\usepackage{mathrsfs}
\usepackage{color}
\usepackage{bm}% bold mathsp
\usepackage{xspace}
\usepackage{epstopdf}
\usepackage{longtable}
\usepackage{amsmath}    % break muiltiline equation into pages
\allowdisplaybreaks[4]  % 1 ~ 4 pages 

\makeatother

\begin{document}
\title{Anti-commutative dynamical magneto-electric response in certain solid
state materials}
\date{\today}
\author{Maoyuan Wang}
\email{mywang6694@pku.edu.cn}

\affiliation{International Center for Quantum Materials, School of Physics, Peking
University, Beijing 100871, China}
\author{Haiwen Liu}
\affiliation{Center for Advanced Quantum Studies, Department of Physics, Beijing
Normal University, Beijing 100875, China}
\author{X. C. Xie}
\affiliation{International Center for Quantum Materials, School of Physics, Peking
University, Beijing 100871, China}
\affiliation{Beijing Academy of Quantum Information Sciences, Beijing 100193, China}
\affiliation{CAS Center for Excellence in Topological Quantum Computation, University
of Chinese Academy of Sciences, Beijing 100871, China}
\begin{abstract}
Axion field induced topological magneto-electric response has attracted
lots of attentions since it was first proposed by Qi \textit{et al.}
in 2008. Here we find a new type of anti-commutative magneto-electric
response $\beta^{\xi}(\omega)$, which can induce a dynamical magneto-electric
current driven by a time-varying magnetic field. Unlike the Chern-Simons
Axion term, this magneto-electric response term is gauge-independent
and non-quantized, and manifests in the systems breaking the symmetries
of the time-reversal, inversion and mirror. In particular, we propose
the antiferromagnetic material Mn$_{2}$Bi$_{2}$Te$_{5}$ as a material
candidate to observe dynamical magneto-electric current, in which
a large magneto-electric response term $\beta^{\xi}(\omega)$ originates
from band inversion.
\end{abstract}
\maketitle

\paragraph*{\textcolor{blue}{Introduction.}}

Magneto-electric response has been discovered more than a century
ago, which describes that an electric field can induce the magnetization
or a magnetic field can induce the polarization in certain materials\citep{fiebig2005revival,spaldin2005renaissance,PhysRevB.82.245118,gao2019semiclassical}.
With magneto-electric response, plenty of relevant novel effects have
been proposed and observed, such as negative magnetoresistance\citep{PhysRevB.88.104412},
chiral magnetic Effect\citep{RN41}, gyrotropic magnetic effect\citep{RN61,PhysRevB.97.035158},
magneto-optical effects\citep{mansuripur1995physical,antonov2004electronic,feng2020topological}. 

Recently, topological magneto-electric response has been discussed
with an effective action $S_{\theta}=\left(\frac{\boldsymbol{e}^{2}}{h}\right)\left(\frac{\theta}{2\pi}\right)\int d^{3}xdt\mathbf{E}\cdot\mathbf{B}$
\citep{RN35,RN6}, similar to Axion in the Standard Model of particle
physics, where $\theta=\pi$ or $\theta=0$ represents the topological
nontrivial or trivial term\citep{RN35,RN30,RN51,RN28,RN17,RN29,RN63,RN38,RN22}.
Generally, topological insulators with the time-reversal symmetry
$\mathscr{T}$ have $\theta=\pi$, while Axion insulators are the
materials without the $\mathscr{T}$ symmetry still maintaining $\theta=\pi$,
protected by other symmetries (such as the $\mathscr{P}$ symmetry
or the mirror symmetry $\mathscr{M}$).  Some novel physical effects
have been proposed in Axion insulators\citep{PhysRevLett.122.256402,PhysRevResearch.2.033342,RN19,RN4,RN48,RN60,liu2021magnetic,PhysRevB.103.L241409},
such as the half-integer anomalous Hall effect on the surface of an
Axion insulator, and the spin-wave excitations induced dynamical axion
effect\citep{RN20,Zhang_2020}. And if the $\mathscr{T}$, $\mathscr{P}$
and $\mathscr{M}$ symmetries are broken, the Chern-Simons Axion term
$\alpha^{CS}=\left(\boldsymbol{e}^{2}/2\pi h\right)\theta$ is not
quantized, and meanwhile certain magneto-electric response terms (such
as the Kubo term) become non-zero\citep{Malashevich_2010,PhysRevB.82.245118,PhysRevB.81.205104,PhysRevB.103.045401,PhysRevB.103.115432}.
General questions arise that whether other type of static or dynamical
magneto-electric response exist in systems without the $\mathscr{T}$,
$\mathscr{P}$ and $\mathscr{M}$ symmetries, and that what microscopic
mechanism determines the new type of magneto-electric response.

In this Letter, we propose a new type of magneto-electric response
term $\beta^{\xi}(\omega)$, originated from the anti-commutative
correlation of the magnetization and the polarization. This term not
only induces the interface magneto-electric responses including the
surface charge and the surface anomalous Hall responses, similar to
the behaviors of Chern-Simons Axion term \citep{RN35,RN6}, but also
gives rise to a new dynamical magneto-electric current response driven
by a time-varying magnetic field $\mathbf{\boldsymbol{j}}^{\beta}=2\mathbf{\beta^{\xi}(\omega)}\partial_{t}\mathbf{\mathbf{B}}$,
with $\beta^{\xi}(\omega)$ denoting the linear response coefficient.
We discuss the origin and required symmetry-breaking terms of this
dynamical magneto-electric response, and propose an effective model
to describe the main features of this phenomenon. We also propose
a small bandgap antiferromagnetic materials Mn$_{2}$Bi$_{2}$Te$_{5}$ as
a possible candidate as well as a feasible experimental setup to observe
the dynamical magneto-electric response.

\paragraph{\textcolor{blue}{Electrodynamics of dynamical magneto-electric effect.}}

\begin{figure}
\centering{}\includegraphics[width=1\columnwidth]{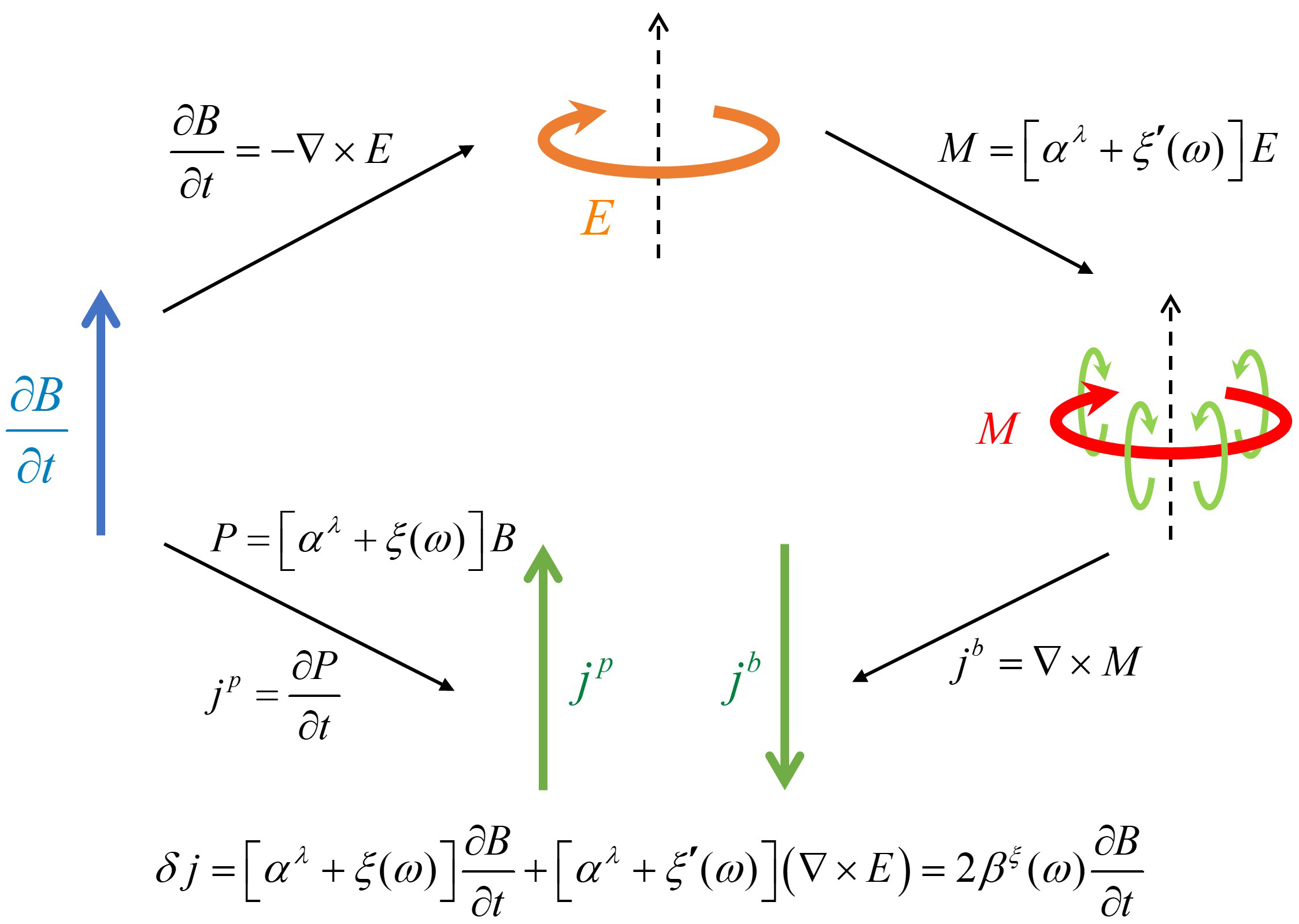}\caption{\textbf{Microscopic process of dynamical magneto-electric effect}.
The schematic includes four parts: 1) polarization current $j^{p}$=$\partial_{t}\mathbf{P}$,
with polarization $\mathbf{P}$ from magneto-electric coupling. 2)
Faraday law $\partial_{t}\mathbf{B}=-\nabla\times\mathbf{E}$. 3)
magneto-electric coupling induced magnetization $\mathbf{M}$. 4)
bound current $j^{b}$=$\nabla\times\mathbf{M}$. The total current
contains the polarization current $j^{p}$ and the bound current $j^{b}$.
Due to the dynamical magneto-electric effect, the time-varying magnetic
field can give rise to total dynamical magneto-electric current $\delta\mathbf{\boldsymbol{j}}^{\beta}=\left[\xi(\omega)-\xi^{\prime}(\omega)\right]\partial\mathbf{\mathbf{B}}/\partial t=2\beta(\omega)\partial\mathbf{\mathbf{B}}/\partial t$
with$\beta(\omega)=\left[\xi(\omega)-\xi^{\prime}(\omega)\right]/2$.
\label{fig:1-1}}
\end{figure}

Considering a system with a longitudinal magneto-electric coupling,
the total Lagrangian can be derived from the linear response theory\footnote{See Supplemental Material for details.}
and reads: 
\begin{eqnarray}
\mathscr{L}(t) & = & \frac{1}{2}\left[\varepsilon_{0}\mathbf{E}^{2}(t)-\frac{1}{\mu_{0}}\mathbf{B}^{2}(t)\right]-\rho\phi+\mathbf{j}\cdot\mathbf{A}\nonumber \\
 &  & -\alpha^{\lambda}(t)\mathbf{E}(t)\cdot\mathbf{B}(t)-\mathbf{E}(t)\int\xi(t,t^{\prime})\mathbf{B}(t^{\prime})dt^{\prime}\nonumber \\
 &  & -\mathbf{B}(t)\int\xi^{\prime}(t,t^{\prime})\mathbf{E}(t^{\prime})dt^{\prime},
\end{eqnarray}
which includes charge density $\rho$, electric potential $\phi$,
current density $\mathbf{j}$, vector potential $\mathbf{A}$.The
last three terms represent the magneto-electric response due to magneto-electric
fields coupled with Bloch electrons in materials. Specifically, $\alpha^{\lambda}(t)$
denotes the simultaneous magneto-electric response, and $\xi(t,t^{\prime})$,
$\xi^{\prime}(t,t^{\prime})$ represent the retarded magneto-electric
response. In the following, we assume that the electric field $\mathbf{E}(t)=\mathbf{E_{0}^{\omega}}e^{-i\omega t}$
and magnetic field $\mathbf{B}(t)=\mathbf{B_{0}^{\omega}}e^{-i\omega t}$
are time-harmonic variables with frequency $\omega$. After Fourier
transform from $t-t^{\prime}$ to $\omega$, the general charge response
and current response can be derived from the Euler-Lagrange equations\footnote{See Supplemental Material for details.}:
\begin{eqnarray}
\delta\rho & = & -\nabla\left[\alpha^{\lambda}+\xi(\omega)\right]\cdot\mathbf{B}(t)
\end{eqnarray}

\begin{eqnarray}
\delta\mathbf{j} & = & \nabla\left[\alpha^{\lambda}+\xi^{\prime}(\omega)\right]\times\mathbf{E}(t)+\partial_{t}\left[\alpha^{\lambda}+\xi(\omega)\right]\mathbf{B}(t)\label{eq:jj}\\
 &  & +2\beta_{\omega}^{\xi}\partial_{t}\mathbf{B}(t)\,.
\end{eqnarray}
Here, the charge and current responses originate from magneto-electric
effect are present, while the ordinary parts with $\rho_{0}\equiv\varepsilon_{0}\nabla\cdot\mathbf{E}(t)$
and $\mathbf{j}_{0}=-\varepsilon_{0}\partial_{t}\mathbf{E}+\frac{1}{\mu_{0}}\nabla\times\mathbf{B}$
are omitted. Comparing with previous studies, the first term of current
responses describes the surface half integer anomalous Hall effect
in an Axion insulator \citep{RN35,RN6}, and the second term describes
the dynamical axion effect \citep{RN20,Zhang_2020}. Both of these
effects should be corrected by the retarded magneto-electric response
$\xi$ and $\xi^{(\prime)}$. The third term represents the dynamical
magneto-electric effect (DME) $\delta\mathbf{\boldsymbol{j}}^{\beta}=2\beta_{\omega}^{\xi}\partial_{t}\mathbf{\mathbf{B}}$
($\beta_{\omega}^{\xi}=\left[\xi(\omega)-\xi^{\prime}(\omega)\right]/2$),
which only depends on the anti-commutative correlation of the magnetization
operator $\hat{\mathbf{M}}$ and the polarization operator $\hat{\mathbf{P}}$
\footnote{See Supplemental Material for details}. The main focus
of this work is the dynamical magneto-electric effect driven by a
time-varying magnetic field denoted by the third term in Eq. \ref{eq:jj}.
As shown in the Fig. \ref{fig:1-1}, for the topological magneto-electric
response term $\alpha^{\lambda}$ (including the case of an Axion
insulator), the polarization current $j^{p}$ and bound current $j^{b}$
cancel each other. Meanwhile, for the retarded magneto-electric responses
$\xi$ and $\xi^{\prime}$, an dynamical magneto-electric current
$\delta\mathbf{\boldsymbol{j}}^{\beta}=2\beta_{\omega}^{\xi}\partial_{t}\mathbf{\mathbf{B}}$
exists, which is a novel effect not studied before.

\paragraph{\textcolor{blue}{General magneto-electric response.}}

\begin{figure*}
\centering{}\includegraphics[width=2\columnwidth]{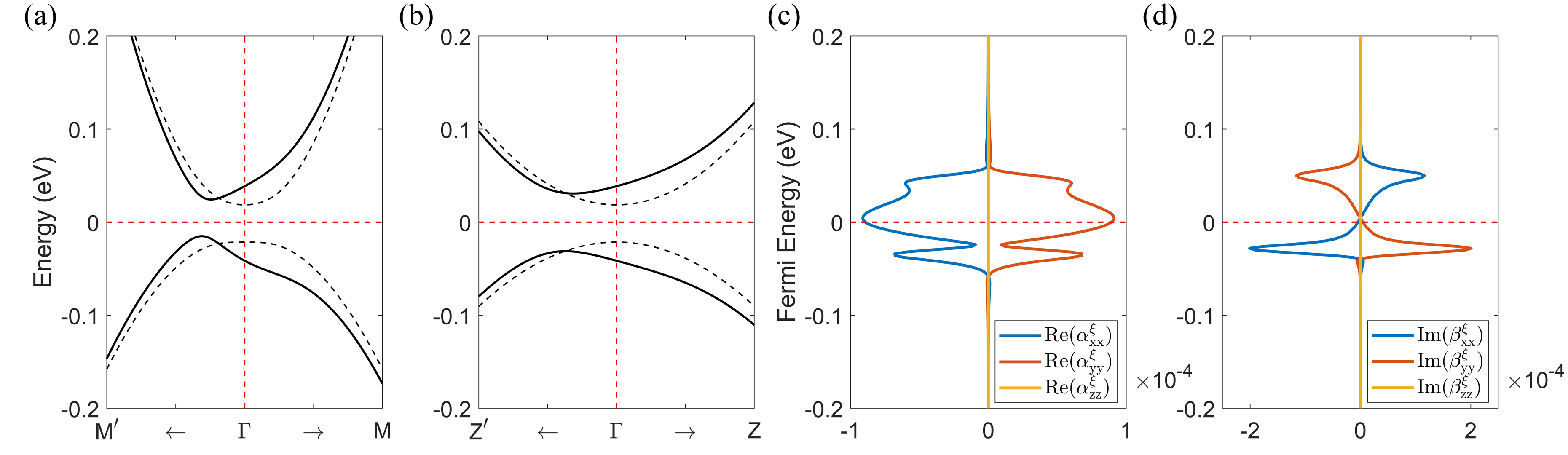}\caption{\textbf{Effective model study of the retarded magneto-electric responses.
}The band-structure in (a) M direction and (b) Z direction. (c) commutative
magneto-electric response $\alpha^{\xi}$ and (d) anti-commutative
magneto-electric response $\beta^{\xi}$ ( in unit of $\boldsymbol{e}^{2}/\hbar$
) ( $m_{1}=m_{2}=m_{3}=0.02$eV in $\delta H$ and $\hbar\omega=0.08$
eV ). \label{fig:2}}
\end{figure*}

The general magneto-electric response can be obtained from the linear
response theory \citep{PhysRevB.99.045121}, including both the simultaneous
magneto-electric response $\alpha^{\lambda}(t,t)$ and the retarded
magneto-electric response $\xi(t,t^{\prime})$ and $\xi^{\prime}(t,t^{\prime})$.
The operator forms of these coefficients read as follows:

\begin{eqnarray}
\alpha^{\lambda}(t,t) & = & -<|\frac{\partial^{2}H}{\partial\mathbf{E}\partial\mathbf{B}}|>_{0}\equiv\alpha^{\lambda}(t)
\end{eqnarray}
\begin{eqnarray}
\xi(t,t^{\prime}) & = & -\frac{i}{\hbar}\Theta(t-t^{\prime})<|\left[\hat{\mathbf{P}}(t),\hat{\mathbf{M}}(t^{\prime})\right]|>_{0}\label{eq:betat}\\
\xi^{\prime}(t,t^{\prime}) & = & -\frac{i}{\hbar}\Theta(t-t^{\prime})<|\left[\hat{\mathbf{M}}(t),\hat{\mathbf{P}}(t^{\prime})\right]|>_{0}\,,
\end{eqnarray}
where $\hat{\mathbf{P}}=-\partial H/\partial\mathbf{E}$ is the polarization
operator, and $\hat{\mathbf{M}}=-\partial H/\partial\mathbf{B}$ is
the orbital magnetization operator. The time variable $t$ in the
brackets represents the measurement time, and $t^{\prime}$ represents
the time of an external electric or a magnetic field. Then, the first
term $\alpha^{\lambda}(t,t)$ is a response function at the same time,
and the Maxwell relations in $\alpha^{\lambda}(t,t)$ make $\mathbf{B}$
and $\mathbf{E}$ commutative in relevant Lagrangian $\mathscr{L}^{\lambda}=-\alpha^{\lambda}(t)\mathbf{E}(t)\cdot\mathbf{B}(t)$.
Thus, $\alpha^{\lambda}(t)$ represents the commutative magneto-electric
response. The second term and the third term describe the magneto-electric
response with different times, $\mathbf{P}(t)=\int\xi(t,t^{\prime})\mathbf{B}(t^{\prime})dt^{\prime},\mathbf{M}(t)=\int\xi^{\prime}(t,t^{\prime})\mathbf{E}(t^{\prime})dt^{\prime}$,
and the related Lagrangian can be written as $\mathscr{L}^{\xi}(t)=-\int\left[\mathbf{E}(t)\xi(t,t^{\prime})\mathbf{B}(t^{\prime})+\mathbf{B}(t)\xi^{\prime}(t,t^{\prime})\mathbf{E}(t^{\prime})\right]dt^{\prime}$,
where $\mathbf{B}$ and $\mathbf{E}$ are not commutative necessarily,
for different times.

Comparing to notations of the optical conductivity in the linear response
theory \citep{PhysRevB.99.045121}, $\alpha^{\lambda}(t)$ term is
the intra-band Drude term, which also includes the Chern-Simons Axion
term $\alpha^{\lambda}=\left(\boldsymbol{e}^{2}/2\pi h\right)\theta$
\citep{Malashevich_2010}. As for the retarded magneto-electric responses
$\xi(t,t^{\prime})$ and $\xi^{\prime}(t,t^{\prime})$, the explicit
formula can be obtained similar to the current-current correlation
in the Kubo formula. To be specific, after Fourier transform from
$t-t^{\prime}$ to frequency $\omega$, these coefficients read: 
\begin{eqnarray}
\xi_{ij}(\omega) & = & \boldsymbol{e}\int\frac{d\boldsymbol{k}}{(2\pi)^{3}}\sum_{n\neq m}\frac{\left(f_{nk}-f_{mk}\right)A_{nm}^{i}M_{mn}^{j}}{\varepsilon_{nk}-\varepsilon_{mk}+\hbar\omega+\mathrm{i}\eta}\label{eq:Qijw}\\
\xi_{ji}^{\prime}(\omega) & = & \boldsymbol{e}\int\frac{d\boldsymbol{k}}{(2\pi)^{3}}\sum_{n\neq m}\frac{\left(f_{nk}-f_{mk}\right)M_{nm}^{j}A_{mn}^{i}}{\varepsilon_{nk}-\varepsilon_{mk}+\hbar\omega+\mathrm{i}\eta}
\end{eqnarray}
where $A_{mn}^{i}=iv_{mn}^{i}/\left(\varepsilon_{nk}-\varepsilon_{mk}\right)$
and $M_{nm}^{j}=\frac{e}{2}\sum_{l\neq m}\left(\boldsymbol{v}_{nl}\times\boldsymbol{A}_{lm}\right)^{j}$
are the inter-band elements of the Berry connection and the orbital
magnetization in the basis of the eigen-functions with the eigen-energies
$\varepsilon_{nk}$\citep{gao2019semiclassical}. In the following,
we mainly consider the longitudinal retarded magneto-electric coefficients
$\xi_{ii}(\omega)$ and $\xi_{ii}^{\prime}(\omega)$, and the transverse
part is beyond the scope of this work. For convenience, one can separate
the commutative part $\alpha_{ii}^{\xi}(\omega)$ and the anti-commutative
part $\beta_{ii}^{\xi}(\omega)$ as follows:
\begin{eqnarray}
 &  & \alpha_{ii}^{\xi}(\omega)=\left[\xi_{ii}(\omega)+\xi_{ii}^{\prime}(\omega)\right]/2\label{eq:aiialpha}\\
 & = & \boldsymbol{e}\int\frac{d\boldsymbol{k}}{(2\pi)^{3}}\sum_{n\neq m}\frac{\left(-i\right)\left(f_{nk}-f_{mk}\right)v_{nm}^{i}M_{mn}^{i}}{\left(\varepsilon_{nk}-\varepsilon_{mk}\right)^{2}-\left(\hbar\omega\right)^{2}}\nonumber 
\end{eqnarray}

\begin{eqnarray}
 &  & \beta_{ii}^{\xi}(\omega)=\left[\xi_{ii}(\omega)-\xi_{ii}^{\prime}(\omega)\right]/2\label{eq:beta}\\
 & = & \boldsymbol{e}\int\frac{d\boldsymbol{k}}{(2\pi)^{3}}\sum_{n\neq m}\frac{-\hbar\omega\left(f_{nk}-f_{mk}\right)A_{nm}^{i}M_{mn}^{i}}{\left(\varepsilon_{nk}-\varepsilon_{mk}\right)^{2}-\left(\hbar\omega\right)^{2}}\,,\nonumber 
\end{eqnarray}
where one can see that $\alpha_{ii}^{\xi}(\omega)$ is real while
$\beta_{ii}^{\xi}(\omega)$ is purely imaginary \footnote{The zero-frequency limit of the commutative part $\alpha^{\xi}(\omega)$
has been discussed in Ref.\citep{Malashevich_2010,PhysRevB.82.245118,PhysRevB.81.205104,PhysRevB.103.045401,PhysRevB.103.115432},
while we focus on anti-commutative part $\beta^{\xi}(\omega)$ here.
When the Chern-Simons Axion term is quantized, at least one of these
symmetries is preserved, therefore $\alpha_{ii}^{\xi}(\omega)=0$
and $\beta_{ii}^{\xi}(\omega)=0$. And see Supplemental Material for
details.}.The anti-commutative magneto-electric coefficient $\beta_{\omega}^{\xi}$
can give rise to new kind of magneto-electric response, which manifests
in systems without the time-reversal symmetry, the inversion symmetry
and the mirror symmetry.

\paragraph*{\textcolor{blue}{A simple effective model.}}

As shown in Eq.\ref{eq:Qijw}, $\xi(\omega)$ is an inter-band gauge-independent
term (similar to $\alpha_{ii}^{\xi}(\omega)$ and $\beta_{ii}^{\xi}(\omega)$),
which is nonzero if the time-reversal symmetry $\mathscr{T}$ and
the inversion symmetry $\mathscr{P}$ of the system are broken, unlike
the Chern-Simons Axion term . Besides, other possible symmetries such
as the mirror symmetry $\mathscr{M}_{x/y/z}$ should also be broken. 

Here we would like to start from a well-known simple model to discuss
the non-zero anti-commutative magneto-electric coefficient $\beta$.
We adopt the Bi$_{2}$Se$_{3}$ effective model discussed in Ref.\citep{zhang2009topological,RN20}
with the basis of bonding and anti-bonding states of the p$_{z}$
orbitals :
\begin{eqnarray}
H & = & H_{0}+\delta H
\end{eqnarray}

\begin{eqnarray}
H_{0} & = & \epsilon_{0}(k)+\sum_{a=1}^{5}d_{a}(k)\Gamma^{a}
\end{eqnarray}

\begin{eqnarray}
\delta H & = & \sum_{i}^{1,2,3,5}m_{i}\Gamma^{i}\,,
\end{eqnarray}
where $\epsilon_{0}(k)=C+2D_{1}+4D_{2}-2D_{1}\mathrm{cos}k_{z}-2D_{2}(\mathrm{cos}k_{x}+\mathrm{cos}k_{y})$,
$d_{1,2,3,4,5}=(A_{2}\mathrm{sin}k_{x},A_{2}\mathrm{sin}k_{y},A_{1}\mathrm{sin}k_{z},M(k),0)$,
$M(k)=M-2B_{1}-4B_{2}+2B_{1}\mathrm{cos}k_{z}+2B_{2}(\mathrm{cos}k_{x}+\mathrm{cos}k_{y})$
and the Dirac matrixes $\Gamma^{1,2,3,4,5}=(\sigma_{x}\otimes s_{x},\sigma_{x}\otimes s_{y},\sigma_{y}\otimes s_{0},\sigma_{z}\otimes s_{0},\sigma_{x}\otimes s_{z})$
in the basis of ($\left|P1_{z}^{+},\uparrow\right\rangle $, $\left|P1_{z}^{+},\downarrow\right\rangle $,
$\left|P2_{z}^{-},\uparrow\right\rangle $, $\left|P2_{z}^{-},\downarrow\right\rangle $).
Here we choose the parameters ultilized previously by Zhang \textit{et
al.} with $A_{1}=0.4$, $A_{2}=0.8$, $B_{1}=2$, $B_{2}=11.32$,
$C=-0.0014$, $D_{1}=0.26$, $D_{2}=3.92$, $M=0.02$\citep{zhang2009topological,RN20}.
$H_{0}$ preserves both the time-reversal symmetry $\mathscr{T}$
and the inversion symmetry $\mathscr{P}$, while $\delta H$ breaks
$\mathscr{T}$ and $\mathscr{P}$ but preserves the combinational
symmetry $\mathscr{\mathscr{P}T}$. Also, $\delta H$ breaks the mirror
symmetry $\mathscr{M}_{x/y/z}$, which may induce a non-zero anti-commutative
magneto-electric response $\beta$.

As we can see in Fig.\ref{fig:2} (a) \& (b), the band structure of
$H_{0}$ ( black dashed lines ) is an even function of the momentum
k because of the $\mathscr{P}$ , $\mathscr{T}$ and the mirror symmetry.
When the $\delta H$ is considered, these symmetries are broken, resulting
in asymmetric band structure ( solid lines in Fig.\ref{fig:2} (a)
\& (b) ). The symmetry breaking terms are associated with the factors
of $\Gamma^{1,2,3}$ in the Hamiltonian $H$. In $H_{0}$, $d_{1,2,3}=(A_{2}\mathrm{sin}k_{x},A_{2}\mathrm{sin}k_{y},A_{1}\mathrm{sin}k_{z})$
are odd in $k_{x}$, $k_{y}$ and $k_{z}$ , respectively. However,
$m_{1,2,3}$ in $\delta H$ are $\boldsymbol{k}$-independant constants.
Thus, $m_{1,2,3}$ in $\delta H$ break the symmetries of $\mathscr{P}$
, $\mathscr{T}$ and mirror $\mathscr{M}_{x/y/z}$. Meanwhile, non-zero
retarded magneto-electric response $\alpha_{ii}^{\xi}(\omega)$ and
$\beta_{ii}^{\xi}(\omega)$ appear, as shown in Fig.\ref{fig:2} (c)
\& (d). Importantly, $\beta_{xx}^{\xi}$ and $\beta_{yy}^{\xi}$ have
the peak values around the band edges, while decrease to zero when
Fermi energy locates in the band gap. In retrospect, the values of
$\alpha_{xx}^{\xi}$ and $\alpha_{yy}^{\xi}$ shows peak values when
Fermi energy locates in the band gap, and decrease when Fermi surface
increases. The prominent values of $\beta_{xx}^{\xi}$ and $\beta_{yy}^{\xi}$
around the band edge can be explained from the remarkable asymmetric
band structure around the band edge as shown in Fig.\ref{fig:2} (a)
\& (b). With increasing the Fermi energy, $H_{0}$ becomes dominant
and $\alpha_{ii}^{\xi}(\omega)$ and $\beta_{ii}^{\xi}(\omega)$ decrease
as shown in Fig.\ref{fig:2} (c) \& (d).
\begin{figure*}
\centering{}\includegraphics[width=1.8\columnwidth]{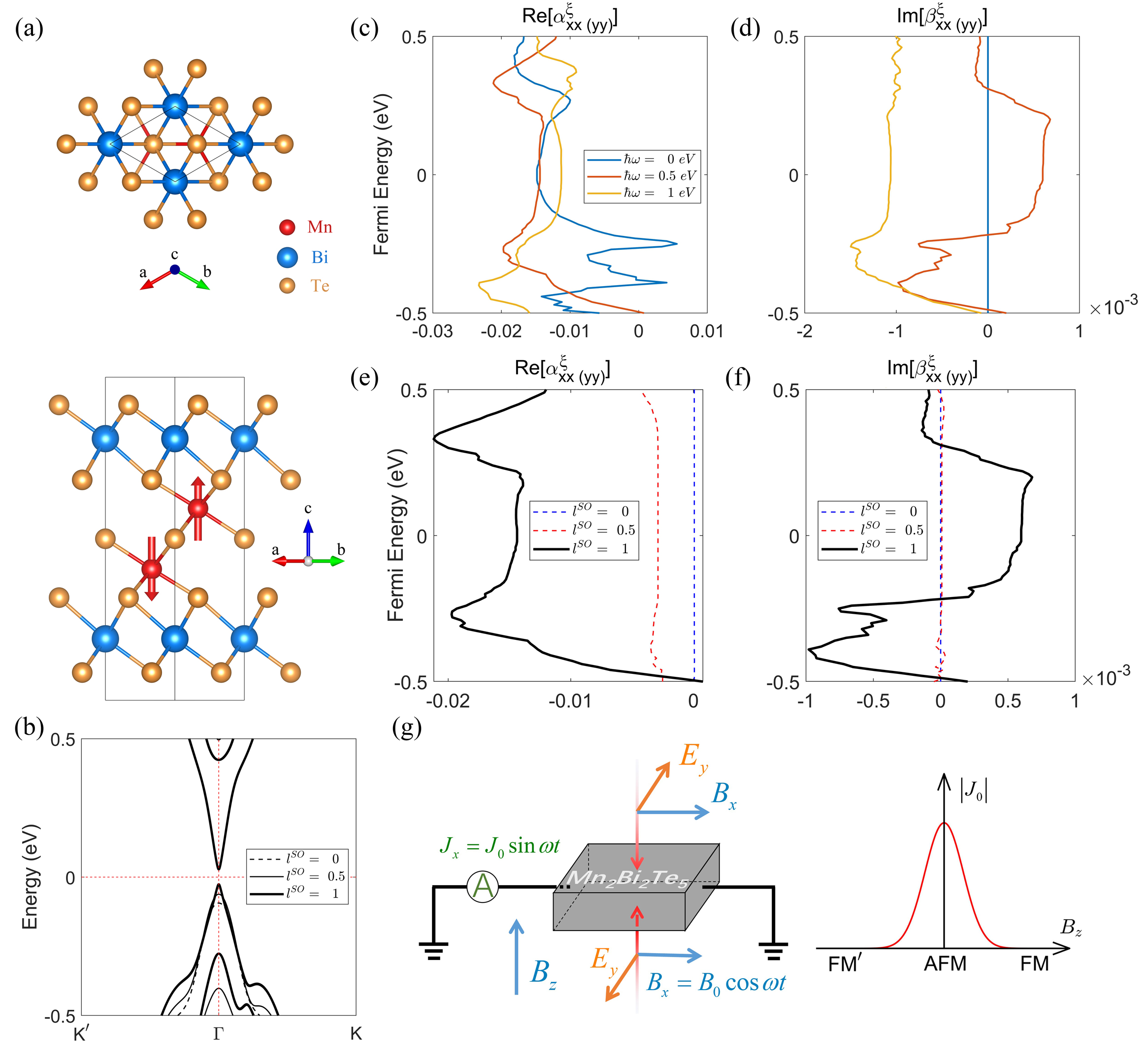}\caption{\textbf{Result of} \textbf{Mn$_{2}$Bi$_{2}$Te$_{5}$}: (a) Crystal
structure and (b) the band-structures with different SOC strength
$l^{SO}$ of Mn$_{2}$Bi$_{2}$Te$_{5}$. (c) - (d) magneto-electric
response $\alpha_{xx(yy)}^{\xi}(\omega)$ and $\beta_{xx(yy)}^{\xi}(\omega)$
( in unit of $\boldsymbol{e}^{2}/\hbar$ ) with different energy $\hbar\omega$
( $l^{SO}$ = 1 ). (e) - (f) $\alpha_{xx(yy)}^{\xi}(\omega)$ and
$\beta_{xx(yy)}^{\xi}(\omega)$ ( in unit of $\boldsymbol{e}^{2}/\hbar$
) with SOC strength $l^{SO}$( $\hbar\omega$ = 0.5 eV ). The effective
SOC strength $\widetilde{\lambda^{SO}}$ satisfies $\widetilde{\lambda^{SO}}=l^{SO}*\lambda^{SO}$,
and $l^{SO}=1$ represents the case of original SOC strength $\lambda^{SO}$.
Here $\alpha(\beta)_{xx}^{\xi}$ is equal to $\alpha(\beta)_{yy}^{\xi}$
due to $\mathscr{C}_{3z}$ symmetry. (g) An experiment setup for observation
of DME. \label{fig:3}}
\end{figure*}

\paragraph*{\textcolor{blue}{A material candidate.}}

Based on the aforementioned model study of the dynamical magneto-electric
response $\beta^{\xi}(\omega)$, the system with nonzero $\beta^{\xi}(\omega)$
needs to break the $\mathscr{P}$ , $\mathscr{T}$ and the mirror
symmetry, and consequently possesses asymmetric band structure. Here
we choose the ternary chalcogenide material Mn$_{2}$Bi$_{2}$Te$_{5}$
as an example. The space group of nonmagnetic Mn$_{2}$Bi$_{2}$Te$_{5}$
is $P\bar{3}m1$ ( No. 164 ) with symmetric operators $\mathscr{P}$
, $\mathscr{T}$, $\mathscr{M}_{x}$, $\mathscr{C}_{3z}$ and $\mathscr{C}_{2x}$
(the crystal structure of Mn$_{2}$Bi$_{2}$Te$_{5}$ is shown in
Fig.\ref{fig:3} (a) )\citep{Zhang_2020,PhysRevB.104.054421}\footnote{See Supplemental Material for details.}
. When the antiferromagnetic order in 001 direction is considered,
a non-zero $\beta^{\xi}(\omega)$ exists in this system. In Fig.\ref{fig:3}
(b), we plot the band structures of Mn$_{2}$Bi$_{2}$Te$_{5}$ with
different values of spin orbital coupling (SOC) strength $\widetilde{\lambda^{SO}}=l^{SO}*\lambda^{SO}$,
where $l^{SO}=1$ represents the case of original SOC strength $\lambda^{SO}$.
When $l^{SO}=0$, the band is symmetric. As the SOC strength increases,
symmetry breaking of $\mathscr{M}_{x}$ in orbital space occurs due
to the SOC effect, resulting in an asymmetric band structure. Meanwhile,
with increasing of the SOC strength, the values of $\alpha_{ii}^{\xi}(\omega)$
and $\beta_{ii}^{\xi}(\omega)$ become larger. As we can see in Fig.\ref{fig:3}
(b), the $\boldsymbol{k}$-path of $\Gamma$ - $\mathrm{K}$ and $\Gamma$
- $\mathrm{K^{\prime}}$ show obvious differences from -0.5 eV to
-0.4 eV, resulting in rapidly changing of $\alpha_{ii}^{\xi}(\omega)$
and $\beta_{ii}^{\xi}(\omega)$ in Fig.\ref{fig:3} (c) \& (d). When
the chosen Fermi Energy is increased to around zero, since the band
asymmetry becomes less obvious, the change of $\alpha_{ii}^{\xi}(\omega)$
and $\beta_{ii}^{\xi}(\omega)$ become slow, and show plateau in the
gap. Besides, the values of $\alpha_{ii}^{\xi}(\omega)$ and $\beta_{ii}^{\xi}(\omega)$
in the gap depend on the system parameters such as $l^{SO}$, as shown
in Fig.\ref{fig:3} (e) \& (f). Moreover, with increasing of SOC's
strength $l^{SO}$, the energy bands become more asymmetric, and the
values of $\alpha_{ii}^{\xi}(\omega)$'s and $\beta_{ii}^{\xi}(\omega)$'s
are getting larger, as shown in Fig.\ref{fig:3} (e) \& (f)\footnote{The values of $\alpha_{ii}^{\xi}(\omega)$'s and $\beta_{ii}^{\xi}(\omega)$
show the asymmetric levels of the occupied bands when the Fermi Energy
is in the bandgap, instead of quatization as the Chern-Simons Axion
term with symmetries protected in Axion insulators.}. According to Ref. \citep{Zhang_2020}, the band structures of Mn$_{2}$Bi$_{2}$Te$_{5}$
are gapless around $l^{SO}=0.9$, and band inversion shows up at larger
$l^{SO}$, which indicates that $\alpha_{ii}^{\xi}(\omega)$ and $\beta_{ii}^{\xi}(\omega)$
can be enhanced in topological materials with band inversion. Specifically,
if $l^{SO}=0$, the inter-band matrix elements is small, such as $A_{nm}^{i}$
in Eq. \ref{eq:beta}, due to two bands near Fermi energy are from
different atoms (Bi and Te). When $l^{SO}>0.9$\citep{Zhang_2020},
the band inversion shows up, which make P$_{z}$ orbital of Bi and
Te mixing, which enhances inter-band matrix elements in Eq. \ref{eq:beta},
and gives rise to a large magneto-electric response term $\beta^{\xi}(\omega)$.
Also, as one can see in Eq. \ref{eq:aiialpha} \& \ref{eq:beta},
$\hbar\omega$ is comparable with the band gap, so the materials with
a small band gap are better for observing the dynamical magneto-electric
response.

\paragraph*{\textcolor{blue}{An experiment design.}}

Finally, we propose an experimental setup to detect the DME current
$\delta\mathbf{\boldsymbol{j}}^{\beta}=2\mathbf{\beta}\partial_{t}\mathbf{\mathbf{B}}$.
In Fig.\ref{fig:3} (g), by adding an alternating magnetic field $\mathrm{B}_{x}$
in the x-direction, one can measure an alternating current in the
same direction. The magnetic field $\mathrm{B}_{x}$ can be obtained
from two light fields with different transmission directions, such
the electric fields of which will cancel each other out (The thickness
of the system should be smaller than the wavelength of the light fields).
According to Ref. \citep{Zhang_2020}, the 001-antiferromagntic state
is the ground state of Mn$_{2}$Bi$_{2}$Te$_{5}$, so we can observe
a non-zero alternating current $J_{x}$ induced by alternating magnetic
fields $\mathrm{B}_{x}$. Furthermore, if we add a large external
static magnetic fields $\mathrm{B}_{z}$ in z-direction and shift
Mn$_{2}$Bi$_{2}$Te$_{5}$ to FM state, the inversion symmetry $\mathscr{P}$
will become preserved and the anti-commutative magneto-electric response
$\beta_{ii}^{\xi}(\omega)$ will vanish. Consequently, there will
be no alternating current $J_{x}$ at a large external static magnetic
field $\mathrm{B}_{z}$, as shown in Fig.\ref{fig:3} (g). For example,
if we have an experimental sample of cross-section size 1 $\mu$m$^{2}$
in x-direction with $\beta_{xx}^{\xi}\sim0.001\left(\boldsymbol{e}^{2}/\hbar\right)$,
and use 1T alternating magnetic field with 1 eV$/h$ frequency, we
can detect an alternating current response of $\sim$ 100 $\mu$A.

\paragraph*{\textcolor{blue}{Conclusion. }}

In summary, we find an anti-commutative magneto-electric response
term $\beta^{\xi}(\omega)$, which can give rise to DME represented
by $\delta\mathbf{\boldsymbol{j}}^{\beta}=2\mathbf{\beta}\partial_{t}\mathbf{\mathbf{B}}$.
This novel magneto-electric response originates from the retarded
magneto-electric response, and the microscopic coefficients are obtained
from the linear response theory. Moreover, the characteristics of
this DME are analyzed by a detailed four-band model, and we find that
a non-zero $\beta^{\xi}(\omega)$ term exists in the systems without
the time-reversal, inversion and the mirror symmetries. The values
of $\beta^{\xi}(\omega)$ can be increased by raising the value of
SOC, which gives rise to a prominent asymmetric band structure. Finally,
we predict Mn$_{2}$Bi$_{2}$Te$_{5}$ as a material candidate and
propose an experimental setup to observe the DME current, for the
band inversion of Mn$_{2}$Bi$_{2}$Te$_{5}$ enhances magneto-electric
response term $\beta^{\xi}(\omega)$.

Maoyuan Wang thanks Jin Cao for helpful discussion. This work was
supported by the Strategic Priority Research Program of Chinese Academy
of Sciences (Grant No. XDB28000000), the National Basic Research Program
of China (Grants No. 2015CB921102, and No. 2017YFA0303301), the National
Natural Science Foundation of China (Grant No. 12022407), and the
China Postdoctoral Science Foundation (Grant No. 2021M700255).

\bibliographystyle{unsrt}

\appendix
\begin{widetext}
\setcounter{figure}{0}
\setcounter{equation}{0}
\renewcommand{\thefigure}{S\arabic{figure}}
\renewcommand{\theequation}{S\arabic{equation}}
\newpage

\section*{Supplemental Materials for \textquotedblleft Anti-commutative dynamical
magneto-electric response in certain solid state materials\textquotedblright}

\subsection{Formula derivation of Euler-Lagrange equations\label{subsec:S1}}

We starts from the Lagrangian with retarded longitudinal magneto-electric
coupling

\begin{eqnarray}
L(\mathbf{r},t) & = & \frac{1}{2}\left[\varepsilon_{0}\mathbf{E}^{2}(\mathbf{r},t)-\frac{1}{\mu_{0}}\mathbf{B}^{2}(\mathbf{r},t)\right]-\rho\phi+\mathbf{j}\cdot\mathbf{A}-\alpha^{\lambda}(\mathbf{r})\mathbf{E}(\mathbf{r},t)\cdot\mathbf{B}(\mathbf{r},t)\nonumber \\
 &  & -\mathbf{E}(t)\int\xi(t,t^{\prime})\mathbf{B}(t^{\prime})dt^{\prime}-\mathbf{B}(t)\int\xi^{\prime}(t,t^{\prime})\mathbf{E}(t^{\prime})dt^{\prime}
\end{eqnarray}
which includes charge density $\rho$, electric potential $\phi$,
current density $\mathbf{j}$, vector potential $\mathbf{A}$, simultaneous
magneto-electric response $\alpha^{\lambda}(t)$ and retarded magneto-electric
response $\xi(t,t^{\prime})$, $\xi^{\prime}(t,t^{\prime})$ . In
the following, the space $\mathbf{r}$ dependence of the field are
not shown for a shorted formula.Combing with relations from Maxwell
equations

\begin{eqnarray}
\mathbf{B}(t) & = & \nabla\times\mathbf{A}(t)\\
\mathbf{E}(t) & = & -\nabla\phi(t)-\dot{\mathbf{A}}(t),
\end{eqnarray}
one can obtain

\begin{eqnarray}
L(t) & = & \frac{1}{2}\sum_{i}\varepsilon_{0}\left[\left(\partial_{i}\phi\right)^{2}+\left(\partial_{t}\mathbf{\mathrm{A}}_{i}\right)^{2}+2\partial_{t}\mathbf{\mathbf{\mathrm{A}}}_{i}\cdot\partial_{i}\phi\right]-\frac{1}{\mu_{0}}\sum_{i}\left[\left(\nabla\mathbf{\mathrm{A}}_{i}\right)^{2}-\partial_{i}\mathbf{A}\cdot\nabla\mathbf{\mathrm{A}}_{i}\right]+\left(-\rho\phi+\mathbf{j}\cdot\mathbf{A}\right)\\
 &  & -\alpha^{\lambda}\left(-\nabla\phi(t)-\dot{\mathbf{A}}(t)\right)\cdot\left(\nabla\times\mathbf{A}(t)\right)-\left(-\nabla\phi(t)-\dot{\mathbf{A}}(t)\right)\int\xi(t-t^{\prime})\mathbf{B}(t^{\prime})dt^{\prime}-\left(\nabla\times\mathbf{A}(t)\right)\int\xi^{\prime}(t-t^{\prime})\mathbf{E}(t^{\prime})dt^{\prime}.\nonumber 
\end{eqnarray}
Then, using Euler-Lagrange equations for scalar potential,

\begin{eqnarray}
\frac{\partial L}{\partial\phi} & = & \partial_{t}(\frac{\partial L}{\partial\dot{\phi}})+\nabla\cdot\frac{\partial L}{\partial\left(\nabla\phi\right)},
\end{eqnarray}
where

\begin{eqnarray}
\frac{\partial L}{\partial\phi} & = & -\rho\\
\frac{\partial L}{\partial\dot{\phi}} & = & 0\\
\frac{\partial L}{\partial\left(\nabla\phi(t)\right)} & = & -\varepsilon_{0}\mathbf{E}(t)+\alpha^{\lambda}\left(\nabla\times\mathbf{A}(t)\right)+\int\xi(t-t^{\prime})\mathbf{B}(t^{\prime})dt^{\prime}\\
\nabla\cdot\frac{\partial L}{\partial\left(\nabla\phi\right)} & = & -\varepsilon_{0}\nabla\cdot\mathbf{E}(t)+\left[\nabla\cdot\alpha^{\lambda}\right]\cdot\left(\nabla\times\mathbf{A}(t)\right)+\nabla\cdot\int\xi(t-t^{\prime})\mathbf{B}(t^{\prime})dt^{\prime}
\end{eqnarray}
one can obtain

\begin{eqnarray}
\frac{\partial L}{\partial\phi} & = & \partial_{t}(\frac{\partial L}{\partial\dot{\phi}})+\nabla\cdot\frac{\partial L}{\partial\left(\nabla\phi\right)}\\
-\rho & = & -\varepsilon_{0}\nabla\cdot\mathbf{E}(t)+\left[\nabla\cdot\alpha^{\lambda}\right]\cdot\left(\nabla\times\mathbf{A}(t)\right)+\nabla\cdot\int\xi(t-t^{\prime})\mathbf{B}(t^{\prime})dt^{\prime}.
\end{eqnarray}
In general, we know that free charge density $\rho_{0}\equiv\varepsilon_{0}\nabla\cdot\mathbf{E}(t)$,
thus we can get extra charge density response from magnetic field

\begin{eqnarray}
\delta\rho & = & \rho-\rho_{0}\\
 & = & -\left[\nabla\cdot\alpha^{\lambda}\right]\cdot\mathbf{B}(t)-\int\left[\nabla\xi(t-t^{\prime})\right]\cdot\mathbf{B}(t^{\prime})dt^{\prime}.\nonumber 
\end{eqnarray}
Moreover, considering the Euler-Lagrange equations for vector potential

\begin{eqnarray}
\frac{\partial L}{\partial\mathbf{\mathrm{A}}_{k}} & = & \partial_{t}(\frac{\partial L}{\partial\mathbf{\mathrm{\dot{A}}}_{k}})+\nabla\cdot\frac{\partial L}{\partial\left(\nabla\mathbf{\mathrm{A}}_{k}\right)},
\end{eqnarray}
where

\begin{eqnarray}
\frac{\partial L}{\partial\mathbf{\mathrm{A}}_{k}} & = & \mathrm{j}_{k}\\
\frac{\partial L}{\partial\mathbf{\mathrm{\dot{A}}}_{k}} & = & \varepsilon_{0}\left(\mathbf{\mathrm{\dot{A}}}_{k}+\partial_{k}\phi\right)+\partial_{t}\left\{ \alpha^{\lambda}\left[\mathbf{B}(t)\right]_{k}\right\} +\partial_{t}\left\{ \int\xi(t-t^{\prime})\left[\mathbf{B}(t^{\prime})\right]_{k}dt^{\prime}\right\} \\
\nabla\cdot\frac{\partial L}{\partial\left(\nabla\mathbf{\mathrm{A}}_{k}\right)} & = & -\nabla\cdot\varepsilon_{0}\left(\nabla\mathbf{\mathrm{A}}_{k}-\partial_{k}\mathbf{A}\right)+\left\{ \nabla\times\left[\alpha^{\lambda}\mathbf{E}(t)\right]\right\} _{k}+\left\{ \nabla\times\int\xi^{\prime}(t-t^{\prime})\mathbf{E}(t^{\prime})dt^{\prime}\right\} _{k},
\end{eqnarray}
one can obtain 

\begin{eqnarray}
\mathbf{j} & = & -\varepsilon_{0}\partial_{t}\mathbf{E}+\frac{1}{\mu_{0}}\nabla\times\mathbf{B}+\partial_{t}\left[\alpha^{\lambda}\mathbf{B}(t)\right]+\partial_{t}\left[\int\xi(t-t^{\prime})\mathbf{B}(t^{\prime})dt^{\prime}\right]\nonumber \\
 &  & +\nabla\times\left[\alpha^{\lambda}\mathbf{E}(t)\right]+\nabla\times\int\xi^{\prime}(t-t^{\prime})\mathbf{E}(t^{\prime})dt^{\prime}.
\end{eqnarray}
Furthermore, based on the relation $\mathbf{j_{0}}\equiv-\varepsilon_{0}\partial_{t}\mathbf{E}+\frac{1}{\mu_{0}}\nabla\times\mathbf{B}$,
then one can obtain the current response from magneto-electric response

\begin{eqnarray}
\delta\mathbf{j} & = & \mathbf{j}-\mathbf{j}_{0}\label{eq:dj}\\
 & = & \partial_{t}\left[\alpha^{\lambda}\mathbf{B}(t)\right]+\partial_{t}\left[\int\xi(t-t^{\prime})\mathbf{B}(t^{\prime})dt^{\prime}\right]+\nabla\times\left[\alpha^{\lambda}\mathbf{E}(t)\right]+\nabla\times\int\xi^{\prime}(t-t^{\prime})\mathbf{E}(t^{\prime})dt^{\prime}\nonumber \\
 & = & \partial_{t}\alpha^{\lambda}\mathbf{B}(t)+\nabla\alpha^{\lambda}\times\mathbf{E}(t)+\partial_{t}\left[\int\xi(t-t^{\prime})\mathbf{B}(t^{\prime})dt^{\prime}\right]+\nabla\times\int\xi^{\prime}(t-t^{\prime})\mathbf{E}(t^{\prime})dt^{\prime},\nonumber 
\end{eqnarray}
Assuming the electric field and magnetic field is time-harmonic with
frequency $\omega$, and Fourier transforming response function $\xi(t-t^{\prime})$
to $\xi(\omega^{\prime})$ 

\begin{eqnarray}
\mathbf{B}(t^{\prime}) & = & \mathbf{B_{0}^{\omega}}e^{-i\omega t^{\prime}}\\
\mathbf{E}(t^{\prime}) & = & \mathbf{E_{0}^{\omega}}e^{-i\omega t^{\prime}}\\
\xi(t-t^{\prime}) & = & \frac{1}{2\pi}\int\xi(\omega^{\prime})e^{-i\omega^{\prime}(t-t^{\prime})}d\omega^{\prime},
\end{eqnarray}
one can obtain

\begin{eqnarray}
\delta\rho & = & \rho-\rho_{0}\\
 & = & -\nabla\alpha^{\lambda}\cdot\mathbf{B}(t)-\int\left[\nabla\xi(t-t^{\prime})\right]\cdot\mathbf{B}(t^{\prime})dt^{\prime},\nonumber 
\end{eqnarray}
in which

\begin{eqnarray}
\int\left[\nabla\cdot\xi(t-t^{\prime})\right]\cdot\mathbf{B}(t^{\prime})dt^{\prime} & = & \int\left[\nabla\int\xi(\omega^{\prime})e^{-i\omega^{\prime}(t-t^{\prime})}d\omega^{\prime}\right]\cdot\mathbf{B_{0}^{\omega}}e^{-i\omega t^{\prime}}dt^{\prime}\\
 & = & \frac{\mathbf{B_{0}^{\omega}}}{2\pi}\nabla\int\int\xi(\omega^{\prime})e^{-i(\omega-\omega^{\prime})t^{\prime}}dt^{\prime}d\omega^{\prime}\\
 &  & \nabla\xi(\omega)\cdot\mathbf{B_{0}^{\omega}}e^{-i\omega t},\nonumber 
\end{eqnarray}
thus one can obtain the charge density response 
\begin{eqnarray}
\delta\rho & = & -\nabla\alpha^{\lambda}\cdot\mathbf{B}(t)-\nabla\xi(\omega)\cdot\mathbf{B}(t)\\
 & = & -\nabla\alpha^{\lambda}\cdot\mathbf{B}(t)-\nabla\left(\alpha_{\omega}^{\xi}+\beta_{\omega}^{\xi}\right)\cdot\mathbf{B}(t).\nonumber 
\end{eqnarray}
As for current response from in Eq.\ref{eq:dj}, where

\begin{eqnarray}
\partial_{t}\left[\int\xi(t-t^{\prime})\mathbf{B}(t^{\prime})dt^{\prime}\right] & = & \partial_{t}\left[\int\frac{1}{2\pi}\int\xi(\omega^{\prime})e^{-i\omega^{\prime}(t-t^{\prime})}d\omega^{\prime}\mathbf{B_{0}^{\omega}}e^{-i\omega t^{\prime}}dt^{\prime}\right]\\
 & = & \mathbf{B_{0}^{\omega}}\partial_{t}\left[\int\frac{e^{-i\omega^{\prime}t}}{2\pi}\int\xi(\omega^{\prime})e^{-i(\omega-\omega^{\prime})t^{\prime}}dt^{\prime}d\omega^{\prime}\right]\nonumber \\
 & = & \mathbf{B_{0}^{\omega}}\partial_{t}\left[\int\frac{e^{-i\omega^{\prime}t}}{2\pi}\int\xi(\omega^{\prime})\delta(\omega-\omega^{\prime})d\omega^{\prime}\right]\nonumber \\
 & = & \partial_{t}\left[\mathbf{B_{0}^{\omega}}e^{-i\omega t}\xi(\omega)\right]\nonumber \\
 & = & \partial_{t}\left[\mathbf{B}(t)\xi(\omega)\right],
\end{eqnarray}
one can obtain current response 

\begin{eqnarray}
\delta\mathbf{j} & = & \partial_{t}\alpha^{\lambda}\mathbf{B}(t)+\nabla\alpha^{\lambda}\times\mathbf{E}(t)+\partial_{t}\left[\mathbf{B}(t)\xi(\omega)\right]+\xi^{\prime}(\omega)\nabla\times\mathbf{E}(t)\\
 & = & \partial_{t}\alpha^{\lambda}\mathbf{B}(t)+\nabla\alpha^{\lambda}\times\mathbf{E}(t)+\frac{\partial\left(\alpha_{\omega}^{\xi}+\beta_{\omega}^{\xi}\right)}{\partial t}\mathbf{B}(t)+\mathbf{\nabla}\left(\alpha_{\omega}^{\xi}-\beta_{\omega}^{\xi}\right)\times\mathbf{E}(t)+2\beta_{\omega}^{\xi}\partial_{t}\mathbf{B}(t).\nonumber \\
 & = & \partial_{t}\left(\alpha^{\lambda}+\alpha_{\omega}^{\xi}+\beta_{\omega}^{\xi}\right)\mathbf{B}(t)+\nabla\left(\alpha^{\lambda}+\alpha_{\omega}^{\xi}-\beta_{\omega}^{\xi}\right)\times\mathbf{E}(t)+2\beta_{\omega}^{\xi}\partial_{t}\mathbf{B}(t).\nonumber 
\end{eqnarray}
The current response is the central phenomenological relation for
dynamical magneto-electric effect as shown in Eq. (3) of the main
text. In the following, we give a microscopic derivation for the DME
based on the linear response theory.

\subsection{Linear response theory of dynamical magneto-electric effect.}

For a system $H_{0}$ with perturbations, the total Hamiltonian 
\begin{equation}
H=H_{0}+V_{1}(t)+V_{2}(t)+\cdots,
\end{equation}
where n in $V_{n}$ marked the n-th order perturbation. And the equation
of motion for density operator reads $i\hbar\dot{\rho}\left(t\right)=\left[H,\rho\right]$,
where $\rho_{0}$ is the unperturbed density operator. According to
standard perturbation procedure \citep{PhysRevB.99.045121}, we replace
$\underset{n}{\sum}V_{n}(t)$ and $\rho$ with $\underset{n}{\sum}\lambda^{n}V_{n}(t)$
and $\underset{n}{\sum}\lambda^{n}\rho_{n}(t)$ in the equation of
motion, and separate terms according to the order of $\lambda$, then
we can get the equation of motion for different order of density operator:

\begin{eqnarray}
i\hbar\dot{\rho}_{0} & = & \left[H_{0},\rho_{0}\right]\\
i\hbar\dot{\rho}_{1}\left(t\right) & = & \left[H_{0},\rho_{1}\left(t\right)\right]+\left[V_{1}\left(t\right),\rho_{0}\right].
\end{eqnarray}

Since the density operator here is in Schr�dinger picture, it is convenient
to utilize the interaction picture, $\rho^{I}\left(t\right)=e^{iH_{0}t}\rho(t)e^{-iH_{0}t}$.
The equation of motion for density operator can be simplified as:

\begin{eqnarray}
i\hbar\dot{\rho}_{0}^{I} & = & 0\\
i\hbar\dot{\rho}_{1}^{I}\left(t\right) & = & \left[V_{1}^{I}\left(t\right),\rho_{0}^{I}\left(t\right)\right].
\end{eqnarray}

After integral, the density operator can be simplified as:

\begin{eqnarray}
\rho_{1}^{I}\left(t\right) & = & -\frac{i}{\hbar}\int_{-\infty}^{t}dt^{\prime}\,\left[V_{1}^{I}\left(t^{\prime}\right),\rho_{0}^{I}\right].
\end{eqnarray}

For a response $R$ excited by an excitation $E$, $V$ , we can identify
response function \citep{PhysRevB.99.045121}:

\begin{eqnarray}
\mathscr{F}^{\left(1\right)0}(t,t) & = & Tr\left[\rho_{0}(t)\frac{\partial\hat{R}(t)}{\partial E(t)}\right]\\
 & = & <|\frac{\partial\hat{R}(t)}{\partial E(t)}|>_{0}\nonumber \\
\mathscr{F}^{\left(1\right)1}(t,t^{\prime}) & = & \Theta_{t^{\prime}}^{t}Tr\left[\frac{\delta\rho_{1}^{I}(t)}{\delta V_{1}^{I}(t^{\prime})}\frac{\partial V_{1}^{I}(t^{\prime})}{\partial E(t^{\prime})}\hat{R}(t)\right]\\
 & = & -\frac{i}{\hbar}\Theta_{t^{\prime}}^{t}<|\left[R(t),\frac{\partial V_{1}^{I}(t^{\prime})}{\partial E(t^{\prime})}\right]|>_{0},\nonumber 
\end{eqnarray}
where $\left(1\right)$ in $\mathscr{F}^{\left(1\right)0}(t,t)$ represent
first order, and the second integer represent two different response
functions. $\mathscr{F}^{\left(1\right)0}(t,t)$ is a simultaneous
response function and $\mathscr{F}^{\left(1\right)1}(t,t^{\prime})$
is a retarded response function. For magneto-electric response, $R$
operator could be Polarization $\mathbf{P}$, excitation $E$ could
be electric field $\mathbf{E}$, perturbation $V_{1}=-\mathbf{M}\cdot\mathbf{B}$,
or $R$ operator could be magnetization operator $\mathbf{M}$, excitation
$E$ could be electric field $\mathbf{\mathbf{B}}$, perturbation
$V_{1}=-\mathbf{P}\cdot\mathbf{E}$.

For $\mathscr{F}^{\left(1\right)0}(t,t)$, we identify it as $\lambda(t,t)$,
\begin{eqnarray}
\lambda(t,t) & = & <|\frac{\partial\hat{\mathbf{P}}}{\partial\mathbf{B}}|>_{0}=<|\frac{\partial\hat{\mathbf{M}}}{\partial\mathbf{E}}|>_{0}\nonumber \\
 & = & -<|\frac{\partial^{2}H}{\partial\mathbf{E}\partial\mathbf{B}}|>_{0}\equiv\alpha^{\lambda}\,.
\end{eqnarray}
As for $\mathscr{F}^{\left(1\right)1}(t,t^{\prime})$, we have two
functions:
\begin{eqnarray}
\xi(t,t^{\prime}) & = & -\frac{i}{\hbar}\Theta(t-t^{\prime})<|\left[\hat{\mathbf{P}}(t),\hat{\mathbf{M}}(t^{\prime})\right]|>_{0}\\
\xi^{\prime}(t,t^{\prime}) & = & -\frac{i}{\hbar}\Theta(t-t^{\prime})<|\left[\hat{\mathbf{M}}(t),\hat{\mathbf{P}}(t^{\prime})\right]|>_{0}\,,
\end{eqnarray}
then we Fourier transform from $t-t^{\prime}$ to frequency $\omega$
in the bulk system with near equilibrium approximation:

\begin{eqnarray}
 &  & \xi_{ij}(\omega=\omega_{1})\\
 & = & \frac{1}{2\pi}\iint_{-\infty}^{+\infty}dtdt^{\prime}e^{i\omega t}\xi_{ij}(t,t^{\prime})\nonumber \\
 & = & \frac{-i}{2\pi\hbar}\int\frac{d\boldsymbol{k}}{(2\pi)^{d}}\iint_{-\infty}^{+\infty}dtdt^{\prime}e^{i(\omega t-\omega_{1}t^{\prime})}\Theta(t-t^{\prime})<|\left[\mathrm{P}_{i}\left(t\right),\mathrm{M}_{j}\left(t^{\prime}\right)\right]|>_{0}\nonumber \\
 & = & \frac{-i}{2\pi\hbar}\sum_{l,l^{\prime},m,m^{\prime}}\int\frac{d\boldsymbol{k}}{(2\pi)^{d}}\iint_{-\infty}^{+\infty}dtdt^{\prime}e^{i(\omega t-\omega_{1}t^{\prime})}\mathrm{P}_{ll^{\prime}}^{i}\mathrm{M}_{mm^{\prime}}^{j}\boldsymbol{\Theta(t-t^{\prime})<|\left[c_{l}^{\dagger}\left(t\right)c_{l^{\prime}}\left(t\right),c_{m}^{\dagger}\left(t^{\prime}\right)c_{m^{\prime}}\left(t^{\prime}\right)\right]|>_{0}}\nonumber \\
 & = & \frac{-i}{2\pi\hbar}\sum_{l,l^{\prime},m,m^{\prime}}\int\frac{d\boldsymbol{k}}{(2\pi)^{d}}\iint_{-\infty}^{+\infty}d(t-t^{\prime})dt^{\prime}e^{i\omega(t-t^{\prime})}e^{i(\omega-\omega_{1})t^{\prime}}\times\nonumber \\
 &  & \,\,\,\,\,\,\,\,\,\,\,\,\,\,\,\,\,\,\,\,\,\,\,\,\,\,\mathrm{P}_{ll^{\prime}}^{i}\mathrm{M}_{mm^{\prime}}^{j}\left[\boldsymbol{g_{l^{\prime}m}^{r}(t-t^{\prime})g_{m^{\prime}l}^{<}(t^{\prime}-t)+g_{l^{\prime}m}^{<}(t-t^{\prime})g_{m^{\prime}l}^{a}(t^{\prime}-t)}\right]\nonumber \\
 & = & \frac{-i}{\left(2\pi\right)^{3}\hbar}\sum_{l,l^{\prime},m,m^{\prime}}\int\frac{d\boldsymbol{k}}{(2\pi)^{d}}\iint_{-\infty}^{+\infty}d(t-t^{\prime})dt^{\prime}\iint_{-\infty}^{+\infty}dE_{1}dE_{2}e^{-i(E_{1}-E_{2})(t-t^{\prime})/\hbar}e^{i\omega(t-t^{\prime})}e^{i(\omega-\omega_{1})t^{\prime}}\times\nonumber \\
 &  & \,\,\,\,\,\,\,\,\,\,\,\,\,\,\,\,\,\,\,\,\,\,\,\,\,\,\mathrm{P}_{ll^{\prime}}^{i}\mathrm{M}_{mm^{\prime}}^{j}\left[g_{l^{\prime}m}^{r}(E_{1})g_{m^{\prime}l}^{<}(E_{2})+g_{l^{\prime}m}^{<}(E_{1})g_{m^{\prime}l}^{a}(E_{2})\right]\nonumber \\
 & = & \frac{-i}{2\pi}\sum_{l,l^{\prime},m,m^{\prime}}\iint\frac{d\boldsymbol{k}}{(2\pi)^{d}}dE_{2}\mathrm{P}_{ll^{\prime}}^{i}\mathrm{M}_{mm^{\prime}}^{j}\left[g_{l^{\prime}m}^{r}(E_{2}+\hbar\omega)g_{m^{\prime}l}^{<}(E_{2})+g_{l^{\prime}m}^{<}(E_{2}+\hbar\omega)g_{m^{\prime}l}^{a}(E_{2})\right]\nonumber \\
 & = & i\iint\frac{d\boldsymbol{k}}{(2\pi)^{d}}\frac{dE}{2\pi}f(E)Tr\left(\mathrm{P}^{i}g_{E+\hbar\omega}^{r}\mathrm{M}^{j}g_{E}^{r-a}+g_{E}^{r-a}\mathrm{M}^{j}g_{E-\hbar\omega}^{a}\mathrm{P}^{i}\right)\nonumber \\
 & = & \int\frac{d\boldsymbol{k}}{(2\pi)^{d}}\sum_{n\neq m}\frac{\left(f_{nk}-f_{mk}\right)\mathrm{P}_{nm}^{i}M_{mn}^{j}}{\varepsilon_{nk}-\varepsilon_{mk}+\hbar\omega+\mathrm{i}\eta},\nonumber 
\end{eqnarray}
where $g^{r},$$g^{a}$, $g^{<}$ are retarded, advanced, lesser Green's
functions, $g^{r-a}=g^{r}-g^{a}$, $\mathrm{P}_{nm}^{i}=\boldsymbol{e}r_{nm}^{i}=\boldsymbol{e}<nk|\hat{r}_{i}|mk>=\boldsymbol{e}<nk|i\partial_{k_{i}}|mk>=\boldsymbol{e}<nk|i\partial_{k_{i}}|mk>=\boldsymbol{e}A_{nm}^{i}$
is inter-band polarization, $A_{mn}^{i}=iv_{mn}^{i}/\left(\varepsilon_{nk}-\varepsilon_{mk}\right)$
is inter-band Berry connection, and $M_{nm}^{j}=\frac{\boldsymbol{e}}{2}\sum_{l\neq m}\left(\boldsymbol{v}_{nl}\times\boldsymbol{A}_{lm}\right)^{j}$
is inter-band elements of orbital magnetization\citep{PhysRevB.82.245118,gao2019semiclassical}.

\begin{eqnarray}
\xi_{ij}(\omega) & = & \boldsymbol{e}\int\frac{d\boldsymbol{k}}{(2\pi)^{d}}\sum_{n\neq m}\frac{\left(f_{nk}-f_{mk}\right)A_{nm}^{i}M_{mn}^{j}}{\varepsilon_{nk}-\varepsilon_{mk}+\hbar\omega+\mathrm{i}\eta}\label{eq:xi1}\\
\xi_{ji}^{\prime}(\omega) & = & \boldsymbol{e}\int\frac{d\boldsymbol{k}}{(2\pi)^{d}}\sum_{n\neq m}\frac{\left(f_{nk}-f_{mk}\right)M_{nm}^{j}A_{mn}^{i}}{\varepsilon_{nk}-\varepsilon_{mk}+\hbar\omega+\mathrm{i}\eta}.\label{eq:xi2}
\end{eqnarray}

Furthermore, we can get two diagonal response functions from linear
combination of $\xi(\omega)$ and $\xi^{\prime}(\omega)$,

\begin{eqnarray}
\alpha_{ii}^{\xi} & = & \left[\xi_{ii}(\omega)+\xi_{ii}^{\prime}(\omega)\right]/2\\
 & = & \boldsymbol{e}\int\frac{d\boldsymbol{k}}{(2\pi)^{d}}\sum_{n\neq m}\frac{\left(f_{nk}-f_{mk}\right)A_{nm}^{i}M_{mn}^{i}}{\varepsilon_{nk}-\varepsilon_{mk}+\hbar\omega+\mathrm{i}\eta}\nonumber \\
 &  & +\boldsymbol{e}\int\frac{d\boldsymbol{k}}{(2\pi)^{d}}\sum_{n\neq m}\frac{\left(f_{nk}-f_{mk}\right)M_{nm}^{i}A_{mn}^{i}}{\varepsilon_{nk}-\varepsilon_{mk}+\hbar\omega+\mathrm{i}\eta}\nonumber \\
 & = & \boldsymbol{e}\int\frac{d\boldsymbol{k}}{(2\pi)^{d}}\sum_{n\neq m}\frac{\left(f_{nk}-f_{mk}\right)A_{nm}^{i}M_{mn}^{i}}{\varepsilon_{nk}-\varepsilon_{mk}+\hbar\omega+\mathrm{i}\eta}\nonumber \\
 &  & +\boldsymbol{e}\int\frac{d\boldsymbol{k}}{(2\pi)^{d}}\sum_{n\neq m}\frac{\left(f_{nk}-f_{mk}\right)A_{nm}^{i}M_{mn}^{i}}{\varepsilon_{nk}-\varepsilon_{mk}-\hbar\omega-\mathrm{i}\eta}\nonumber \\
 & = & \boldsymbol{e}\int\frac{d\boldsymbol{k}}{(2\pi)^{d}}\sum_{n\neq m}\frac{\left(f_{nk}-f_{mk}\right)\left(\varepsilon_{nk}-\varepsilon_{mk}\right)A_{nm}^{i}M_{mn}^{i}}{\left[\left(\varepsilon_{nk}-\varepsilon_{mk}\right)^{2}-\left(\hbar\omega+\mathrm{i}\eta\right)^{2}\right]}\nonumber \\
 & = & \boldsymbol{e}\int\frac{d\boldsymbol{k}}{(2\pi)^{d}}\sum_{n\neq m}\frac{\left(-i\right)\left(f_{nk}-f_{mk}\right)v_{nm}^{i}M_{mn}^{i}}{\left[\left(\varepsilon_{nk}-\varepsilon_{mk}\right)^{2}-\left(\hbar\omega+\mathrm{i}\eta\right)^{2}\right]}\nonumber \\
\beta_{ii}^{\xi}(\omega) & = & \left[\xi_{ii}(\omega)-\xi_{ii}^{\prime}(\omega)\right]/2\nonumber \\
 & = & \boldsymbol{e}\int\frac{d\boldsymbol{k}}{(2\pi)^{d}}\sum_{n\neq m}\frac{\left(f_{nk}-f_{mk}\right)A_{nm}^{i}M_{mn}^{i}}{\varepsilon_{nk}-\varepsilon_{mk}+\hbar\omega+\mathrm{i}\eta}\nonumber \\
 &  & -\boldsymbol{e}\int\frac{d\boldsymbol{k}}{(2\pi)^{d}}\sum_{n\neq m}\frac{\left(f_{nk}-f_{mk}\right)M_{nm}^{i}A_{mn}^{i}}{\varepsilon_{nk}-\varepsilon_{mk}+\hbar\omega+\mathrm{i}\eta}\nonumber \\
 & = & \boldsymbol{e}\int\frac{d\boldsymbol{k}}{(2\pi)^{d}}\sum_{n\neq m}\frac{\left(f_{nk}-f_{mk}\right)A_{nm}^{i}M_{mn}^{i}}{\varepsilon_{nk}-\varepsilon_{mk}+\hbar\omega+\mathrm{i}\eta}\nonumber \\
 &  & -\boldsymbol{e}\int\frac{d\boldsymbol{k}}{(2\pi)^{d}}\sum_{n\neq m}\frac{\left(f_{nk}-f_{mk}\right)A_{nm}^{i}M_{mn}^{i}}{\varepsilon_{nk}-\varepsilon_{mk}-\hbar\omega-\mathrm{i}\eta}\nonumber \\
 & = & \boldsymbol{e}\int\frac{d\boldsymbol{k}}{(2\pi)^{d}}\sum_{n\neq m}\frac{-\hbar\omega\left(f_{nk}-f_{mk}\right)A_{nm}^{i}M_{mn}^{i}}{\left[\left(\varepsilon_{nk}-\varepsilon_{mk}\right)^{2}-\left(\hbar\omega+\mathrm{i}\eta\right)^{2}\right]}.\label{eq:betaii}
\end{eqnarray}
These dynamical magneto-electric coefficients represent the main microscopic
results of our work as shown in Eq. (8)-(9) of the main text.

Next, we discuss about gauge-relavant issues. Generally, gauge-dependence
comes from partial differential operator, such as in Berry connection
$A_{mm}^{i}=<mk|i\partial_{k_{i}}|mk>$. If we make a gauge transform
$|\tilde{mk}>=e^{i\psi_{m}(k)}|mk>$, 
\begin{eqnarray}
\tilde{A}_{mm}^{i} & = & <\tilde{mk}|i\partial_{k_{i}}|\tilde{mk}>\\
 & = & <mk|e^{-i\psi_{m}(k)}e^{i\psi_{m}(k)}\left(i\partial_{k_{i}}|mk>\right)+<mk|e^{-i\psi_{m}(k)}\left(i\partial_{k_{i}}e^{i\psi_{m}(k)}\right)|mk>\nonumber \\
 & = & <mk|i\partial_{k_{i}}|mk>-<mk|e^{-i\psi_{m}(k)}e^{i\psi_{m}(k)}\left(\partial_{k_{i}}\psi_{m}(k)\right)|mk>\nonumber \\
 & = & A_{mm}^{i}-\partial_{k_{i}}\psi_{m}(k)\,,\nonumber 
\end{eqnarray}
therefore the (intra-band) Berry connection depend on the phase $\psi(k)$
chose in the gauge. 

However, for the inter-band Berry connection $A_{nm}^{i}=<nk|i\partial_{k_{i}}|mk>\,\left(n\neq m\right)$,
the first term of different gauge is
\begin{eqnarray}
 &  & <nk|e^{-i\psi_{n}(k)}e^{i\psi_{m}(k)}\left(i\partial_{k_{i}}|mk>\right)\\
 & = & e^{i\left[\psi_{m}(k)-\psi_{n}(k)\right]}<nk|i\partial_{k_{i}}|mk>,\,\left(n\neq m\right)\nonumber 
\end{eqnarray}
and the second term is
\begin{eqnarray}
 &  & <nk|e^{-i\psi(k)}e^{i\psi(k)}\left(\partial_{k_{i}}\psi(k)\right)|mk>\nonumber \\
 & = & \left(\partial_{k_{i}}\psi(k)\right)<nk|mk>\nonumber \\
 & = & 0,\,\left(n\neq m\right)
\end{eqnarray}
due to the orthogonality of two states. Therefore, the interband Berry
connection$\tilde{A}_{nm}^{i}=e^{i\left[\psi_{m}(k)-\psi_{n}(k)\right]}A_{nm}^{i}$.
Similarly, interband orbital magnetization of different gauge is 
\begin{eqnarray}
\tilde{M}_{mn}^{j} & = & \frac{\boldsymbol{e}}{2}\sum_{l\neq n}\left(\boldsymbol{\tilde{v}}_{ml}\times\boldsymbol{\tilde{A}}_{ln}\right)^{j}\\
 & = & \frac{\boldsymbol{e}}{2}\sum_{l\neq n}\left(\boldsymbol{v}_{ml}e^{i\left[\psi_{l}(k)-\psi_{m}(k)\right]}\times\boldsymbol{A}_{ln}e^{i\left[\psi_{n}(k)-\psi_{l}(k)\right]}\right)^{j}\nonumber \\
 & = & e^{i\left[\psi_{n}(k)-\psi_{m}(k)\right]}M_{mn}^{j}.\nonumber 
\end{eqnarray}

So, the inter-band Berry connection times interband orbital magnetization
$\tilde{A}_{nm}^{i}\tilde{M}_{mn}^{j}=A_{nm}^{i}M_{mn}^{j}\,\left(n\neq m\right)$
in Eq. \ref{eq:xi1}, Eq. \ref{eq:xi2} and Eq. \ref{eq:betaii} is
gauge independent due to their two extra phases canceling each other,
which explains that the magneto-electric response term $\beta$ is
gauge-independent or gauge invariant.

\end{widetext}

\subsection{Symmetry analyses for magneto-electric coefficients}

Magneto-electric response describes that an electric field can induce
the magnetization $\mathbf{M}=a\mathbf{E}$ with a coefficient 
\begin{eqnarray}
a & = & \alpha-\beta
\end{eqnarray}
 or a magnetic field can induce the polarization $\mathbf{P}=b\mathbf{B}$
with a coefficient 
\begin{eqnarray}
b & = & \alpha+\beta\,.
\end{eqnarray}
Using an inversion symmetry operation, we have 

\begin{eqnarray}
\mathscr{P}\left(\mathbf{E}\right) & = & -\mathbf{E}\\
\mathscr{P}\left(\mathbf{B}\right) & = & \mathbf{B}\\
\mathscr{P}\left(\mathbf{P}\right) & = & -\mathbf{\mathbf{P}}\\
\mathscr{P}\left(\mathbf{M}\right) & = & \mathbf{M},
\end{eqnarray}
and further

\begin{eqnarray}
\mathscr{P}\left(\mathbf{M}\right) & = & \mathscr{P}\left(a\right)\mathscr{P}\left(\mathbf{E}\right)\\
\mathbf{M} & = & -\mathscr{P}\left(a\right)\mathbf{E}
\end{eqnarray}
for example. Combining $\mathbf{M}=a\mathbf{E}$ and $\mathbf{M}=-\mathscr{P}\left(a\right)\mathbf{E}$,
we have $\mathscr{P}\left(a\right)=-a$, which indicates that in the
system with the inversion symmetry, the gauge-independent coefficient
$a=0$ (If $a$ is a gauge-dependent coefficient like Chern Simon
term $\alpha^{CS}=\left(\boldsymbol{e}^{2}/2\pi h\right)\theta$,
where $\theta\rightarrow\theta+2\pi n$ under gauge transform, the
inversion symmetry make $\theta=\pi n$, where $n$ is an integer).
As for coefficient $b$, since $\mathbf{\mathscr{P}\left(P\right)}=-\mathbf{\mathbf{P}}=\mathscr{P}\left(b\right)\mathbf{B}$,
it is the same as $a$ with inversion symmetry. And if both $a$ and
$b$ are zero, both $\alpha$ and $\beta$ are zero too. Generally,
one can consider these coefficients as tensors, and diagonal coefficients
$a_{ii}$, $b_{ii}$ and $\alpha_{ii}$, $\beta_{ii}$ also vanish
in system with inversion symmetry similar to the case shown above.
(offdiagonal magneto-electric coefficients $a_{ij}$ and $b_{ji}$
are not discussed in details in this work).

Similarly, using a time-reversal symmetry $\mathscr{T}$ operation, 

\begin{eqnarray}
\mathscr{T}\left(\mathbf{E}\right) & = & \mathbf{E}\\
\mathscr{T}\left(\mathbf{B}\right) & = & \mathbf{-B}\\
\mathscr{T}\left(\mathbf{P}\right) & = & \mathbf{\mathbf{P}}\\
\mathscr{T}\left(\mathbf{M}\right) & = & \mathbf{-M},
\end{eqnarray}
 it has similar results that $\mathscr{T}\left(a\right)=-a,\mathscr{T}\left(b\right)=-b$.

Besides, mirror symmetry is also important, which usually has three
operators $\mathscr{M}_{x/y/z}$. Using these operators, we have

\begin{eqnarray}
\mathscr{M}_{i}\left(\mathrm{E}_{j}\right) & = & \left(-1\right)^{\delta_{ij}}\mathrm{E}_{j}\\
\mathscr{M}_{i}\left(\mathrm{B}_{j}\right) & = & \left(-1\right)^{1-\delta_{ij}}\mathrm{B}_{j}\\
\mathscr{M}_{i}\left(\mathrm{P}_{j}\right) & = & \left(-1\right)^{\delta_{ij}}\mathrm{P}_{j}\\
\mathscr{M}_{i}\left(\mathrm{M}_{j}\right) & = & \left(-1\right)^{1-\delta_{ij}}\mathrm{M}_{j},
\end{eqnarray}
where index i (j) represents vector component's direction. For example,
if we consider magneto-electric response $\mathrm{M}_{z}=a_{zz}\mathrm{E}_{z}$,
$\mathrm{P}_{z}=b_{zz}\mathrm{B}_{z}$, the mirror operator $\mathscr{M}_{z}$
results in $\mathrm{M}_{z}=\mathscr{M}_{z}\left(a_{zz}\right)\left(-\mathrm{E}_{z}\right)$,
$\left(-\mathrm{P}_{z}\right)=\mathscr{M}_{z}\left(b_{zz}\right)\mathrm{B}_{z}$,
while the mirror operator $\mathscr{M}_{x/y}$ results in $\left(-\mathrm{M}_{z}\right)=\mathscr{M}_{x/y}\left(a_{zz}\right)\mathrm{E}_{z}$,
$\mathrm{\mathrm{P}_{z}}=\mathscr{M}_{z}\left(b_{zz}\right)\left(-\mathrm{B}_{z}\right)$,
which further makes $\mathscr{M}_{x/y/z}\left(a_{zz}\right)=-a_{zz}$,
$\mathscr{M}_{x/y/z}\left(b_{zz}\right)=-b_{zz}$. Therefore, any
mirror symmetry in one system will make gauge-independent diagonal
magneto-electric coefficients $a_{ii}=0,\,b_{ii}=0$ (and further
$\alpha_{ii}=0,\,\beta_{ii}=0$) (offdiagonal magneto-electric coefficients
$a_{ij}$ and $b_{ji}$ can retain one mirror symmetry $\mathscr{M}_{j}$).

For the off-diagonal magneto-electric coefficients of the alpha and
beta tensors $a_{ij}$ and $b_{ji}$,  it is not relevant to our work,
because we study the type of magneto-electric response like $E\cdot B$,
which only involves diagonal magneto-electric coefficients $a_{ii}$
and $b_{ii}$ in the Lagrangian:
\begin{eqnarray}
\mathscr{L}^{ME} & = & \sum_{i}B_{i}a_{ii}E_{i}+E_{i}b_{ii}B_{i}\,.
\end{eqnarray}
The off-diagonal magneto-electric coefficients involve in the magneto-electric
response type like

\begin{eqnarray}
\mathscr{L}^{ME} & = & \sum_{i\neq j}B_{i}a_{ij}E_{j}+E_{j}b_{ji}B_{i}\,,
\end{eqnarray}
which needs systematic study in further work.

\subsection{Calculation details for Mn$_{2}$Bi$_{2}$Te$_{5}$ }

The first-principles calculations for Mn$_{2}$Bi$_{2}$Te$_{5}$
are performed using Vienna \emph{ab initio }simulation package (VASP)
\citep{PhysRevB.54.11169,PhysRevB.59.1758} based on the density function
theory with Perdew-Burke-Ernzerhof (PBE) parameterization of generalized
gradient approximation (GGA)\citep{PhysRevLett.77.3865}. The energy
cutoff of the plane wave basis is set as 300 eV, and the Brillouin
zone is sampled by 12 \texttimes{} 12 \texttimes{} 4 k-mesh. 

The magneto-electric coefficients are calculated based on the Hamiltonian
of maximally localized Wannier functions (MLWF)\citep{mostofi2008wannier90}
with s, d orbitals of Mn atoms, and s, p orbitals of Bi atoms and
Te atoms, with 100 \texttimes{} 100 \texttimes{} 40 k-mesh in the
Brillouin zone. To be more explicit, after MLWF calculations, we have
MLWF Hamiltonian and position operator matrix 
\begin{eqnarray}
H_{ij}(R) & = & <0i|H|jR>\\
r_{ij}^{x/y/z}(R) & = & <0i|r^{x/y/z}|jR>\,,
\end{eqnarray}
where i, j represent MLWF orbitals, R represents cell vector and $r^{x/y/z}$
represents position operator $r^{x}$, $r^{y}$ or $r^{z}$ with different
directions. Using Fourier transforming, we have Hamiltonian and position
operator matrix in k-space

\begin{eqnarray}
H_{ij}^{W}(k) & = & \sum_{R}e^{ikR}H_{ij}(R)\\
r_{ij}^{x/y/z}(k) & = & \sum_{R}e^{ikR}r_{ij}^{x/y/z}(R)\\
 & \equiv & A_{ij}^{x/y/z,W}(k)
\end{eqnarray}
and also velocity operator\citep{PhysRevB.74.195118} 
\begin{eqnarray}
v_{ij}^{x/y/z,W}(k) & = & \partial_{k}H_{ij}^{W}(k)+\mathrm{i}[H^{W}(k),A^{x/y/z,W}(k)]_{ij}\,.
\end{eqnarray}
After diagonalization $E_{k}=U^{\dagger}H^{W}(k)U$, we have eigen
energy $E_{nn}=\varepsilon_{nk}$ and further velocity operator matrix,
berry connection matrix and orbital magnetization matrix in the basis
of eigen vectors
\begin{eqnarray}
v_{mn}^{x/y/z}(k) & = & \left(U^{\dagger}v^{x/y/z,W}(k)U\right)_{mn}\\
A_{mn}^{x/y/z} & = & \frac{\mathrm{i}v_{mn}^{x/y/z}}{\varepsilon_{nk}-\varepsilon_{mk}}\\
M_{nm}^{x/y/z} & = & \frac{\boldsymbol{e}}{2}\sum_{l\neq m}\left(\boldsymbol{v}_{nl}\times\boldsymbol{A}_{lm}\right)^{x/y/z}\,,
\end{eqnarray}
which can be used to calculate magneto-electric coefficients mentioned
before.

\bibliography{Ref}
\end{document}